\documentclass[twocolumn,showpacs,amsmath,amssymb]{revtex4}

\usepackage{graphicx}
\usepackage{dcolumn}
\usepackage{bm}
\usepackage[figuresright]{rotating}
\usepackage{wrapfig}
\usepackage{subfigure}

\def\p{\partial}
\def\f{\frac}
\def\d{{\rm d}}

\begin{document}

\title{Gravity-free hydraulic jumps and metal femtocups}

\author{Rama Govindarajan$^{1,*}$, Manikandan Mathur$^{1,\dag}$, Ratul 
DasGupta$^1$, N.R. Selvi$^2$, Neena Susan John$^2$ and G.U. Kulkarni$^{2,*}$}
\affiliation{1. Engineering Mechanics Unit and 2. Chemistry and Physics of
Materials Unit and DST Unit on Nanoscience, Jawaharlal Nehru Centre for 
Advanced Scientific Research, Jakkur, Bangalore 560064, India. \\
$\dag$ currently at Dept. of Mechanical Engineering, MIT, Cambridge, MA 
02139, USA. 
}

\date{\today}

\begin{abstract}

Hydraulic jumps created by gravity are seen every day in the kitchen sink. 
We show that at small scales a circular hydraulic jump can be created in 
the absence of gravity, by surface tension. The theory is motivated by
our experimental finding of a height discontinuity in spreading submicron 
molten metal droplets created by pulsed-laser ablation. By careful 
control of initial conditions, we show that this leads to solid femtolitre 
cups of gold, silver, copper, niobium and tin. 

\end{abstract}


\maketitle

It has long been observed that water flowing horizontally
can display a discontinuity in height \cite{rayl,watson,tani}. This is the  
{\em hydraulic jump}, seen for example when water from a faucet impinges on the kitchen sink and spreads outwards. 
Gravity is a key ingredient in these well-understood {\em large-scale} hydraulic jumps, as discussed briefly below. 
In this Letter we show, remarkably, that the shallow water equations support 
solutions for a {\em gravity-free} hydraulic jump. The driver here is surface 
tension at the liquid-air (or liquid-vacuum) interface, and jumps may be 
expected to occur when relevant length scales are submicron.
Our theoretical study was prompted by careful 
experiments showing that molten metal droplets impinging on a solid substrate 
display such a jump, solidifying into cup-shaped containers of 
femtolitre capacity. The droplets are created by laser-ablation of a solid 
metal target. Femtocups made of different metals on various substrates
are formed under carefully maintained conditions of 
laser energy and substrate temperature. Outside this narrow range of 
parameters we find what one would normally expect: the droplets solidify 
into lump-shaped structures on the substrate. The ability to make, and
subsequently leach out, femtocups 
at will has potential applications ranging from nanoscale synthetic chemistry 
to single cell biology.

Before describing the experiments and the femtocups, we discuss what causes a
gravity-free hydraulic jump. Consider a steady axisymmetric jet of fluid
of radius $a$ impinging on a solid plate placed normal to the flow. The
density of the surrounding medium is assumed to be negligible. The
fluid then spreads radially outwards within a relatively thin film. The 
dynamics within the film is described by the 
axisymmetric shallow-water equation \cite{30,bohr1,bush} 
\begin{equation}
u\f{\p u}{\p r} + w \f{\p u}{\p z} = \nu \f{\p^2u}{\p z^2} - g h' +
\f{\sigma}{\rho} \f{\d}{\d r}\left[\f{\nabla^2h + (h'^3/r)}{(1 +
h'^2)^{3/2}}\right], 
\label{shallow}
\end{equation}
where $r$ and $z$ are the radial coordinate and the coordinate perpendicular 
to the solid wall respectively, with origin on the solid surface at the 
centre of the impinging jet. The respective velocity components are $u$ and 
$w$. The total height $h$ of the fluid above the surface is a function of 
$r$, and a prime thus denotes a derivative with respect to $r$.  
The parameters in the problem are the acceleration due to 
gravity, $g$, the surface tension coefficient, $\sigma$ for the liquid-air or 
liquid vacuum interface, and the density $\rho$ and the kinematic viscosity 
$\nu$ of the impinging fluid. For incompressible axisymmetric flow the equation 
of continuity, in differential and in global form, reads 
\begin{equation}
\f{\partial u}{\partial r}+\f{u}{r}+\f{\partial w}{\partial z}=0 \quad {\rm
and} \quad 2 \pi \int_0^{h(r)} ru(r,z)dz = Q,
\label{cont}
\end{equation}
where $Q = \pi a^2u_j$ is the steady inlet volumetric flow rate. A
characteristic inlet jet velocity $u_j$ is thus defined.
It is reasonable to assume \cite{tani} a parabolic shape in $z$ for the 
radial velocity 
\begin{equation}
u(r,z)=\zeta(r) (z^{2}-2h(r) z)
\label{parabolic}
\end{equation}
satisfying the no-slip condition at the wall ($z=0$) and the zero shear stress
condition at the free surface ($z=h$). The analysis does not hinge on this
assumption; any reasonable profile shape will give qualitatively the same
results. Using Eq. (\ref{cont}) and the kinematic condition 
$w=$D$h/$D$t=uh^{\prime }$ at $z=h$, the momentum equation (\ref{shallow}) 
integrated over $z$ from $0$ to $h$ reduces after some algebra to
\begin{equation}
b\left( \f{h}{r}+h^{\prime }\right) =\f{2r}{R}+\f{h^{3}r^{2}}{F}
h^{\prime }-\f{r^{2}h^{3}}{W}\f{d}{dr}\left[ \f{\nabla
^{2}h+(h^{\prime 3}/r)}{(1+h^{\prime }{}^{2})^{3/2}}\right], 
\label{final}
\end{equation}
where all lengths are scaled by $a$, and the $O(1)$ positive constant $b = 2/5$ for a parabolic profile. The left-hand side of (\ref{final}) represents 
inertia, and the three terms on the right hand side appear due to viscosity, 
gravity and surface tension respectively. The relative importance of the
inertial term to each of these is quantified respectively by the Reynolds
number $R \equiv u_ja/\nu$ $R$, the Froude number $F \equiv u_j^2/(ga)$, and 
the Weber number $W \equiv \rho u_j^2a /\sigma$. 
In large-scale flows surface tension has been shown \cite{bush} only to 
make a small correction to the location of the jump, so the last term is
unimportant.
This is to be expected, since $F$ in the kitchen sink is of order 
unity, while $W \sim 10-100$. In contrast, consider $u_j \sim 10$
m/s and $a \sim 10^{-7}$m, so $F \sim 10^8$ and $W \sim 10^{-2}$. Here
surface tension determines whether and where a jump will occur, whereas it 
is the {\em gravity} term that may be dropped entirely from the equation. 

In general, a jump occurs if the pressure gradient becomes increasingly 
adverse as the flow proceeds downstream, and attains a magnitude large enough 
to counter the relevant inertial effects. The adverse pressure gradient may 
be created by gravity, or surface tension, or both. 
With gravity alone, Eq. (\ref{final}) reduces to
\begin{equation}
h^{\prime}= \f{2r/R - b h/r}{b - h^3r^2/F}.  
\label{gravity}
\end{equation}
It is seen that if $F$ is finite and $h^3r^2$ is an increasing function of $r$,
the denominator will go to zero at some $r$, i.e., a jump will occur in the framework
of the shallow-water equations \cite{rayl,watson,bohr1,bush}. 
However, its precise location may not coincide with this estimate \cite{bohr1}, and 
in a given experiment the radial extent available may be too small, 
or the inertia too low, for a jump to occur. Decreasing gravity has been 
shown to shift the jump location downstream \cite{28}, consistent with
Eq. (\ref{gravity}). 
Now considering surface tension alone, a crude prediction of the
existence of a jump may be made by assuming the height upstream to be slowly varying
in $r$, i.e., $h^{\prime}<< 1$, and thus setting $h^{\prime\prime}=h^{\prime
\prime\prime}=0$. We may then rewrite Eq. (\ref{final}) as 
\begin{equation}
h^{\prime}\simeq \f{2r/R - b h/r}{b - h^3/W}.  
\label{surface_approx}
\end{equation}
A jump is now possible if $h$ is an increasing function of $r$, which is a
more stringent requirement than in the case of gravity. Note that the second
term in the denominator appears due to {\em radial} spreading, i.e., surface 
tension alone cannot give rise to a one-dimensional jump like a tidal bore.

\begin{figure}
\includegraphics[width=0.25\textwidth]{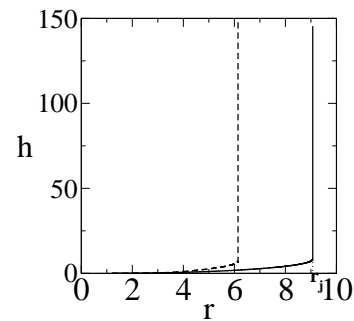}
\vskip3mm
\caption{Typical solutions of Eq. (\ref{final}), with $F=\infty$,
containing a singularity at $r=r_j$. For demonstration, liquid 
properties are taken as those of molten silver ($\rho=5000$ kg m$^{-3}$, 
$\nu=10^{-6}$m/s$^2$ and $\sigma=0.9$Nm$^{-1}$) and $a=5 \mu$m. Solid 
line: $u_j=5$ cm/s; dashed line: $u_j=80$ m/s. Values for molten
tin show similar behavior.}
\label{typical}
\end{figure}

\begin{figure}
\includegraphics[width=0.15\textwidth]{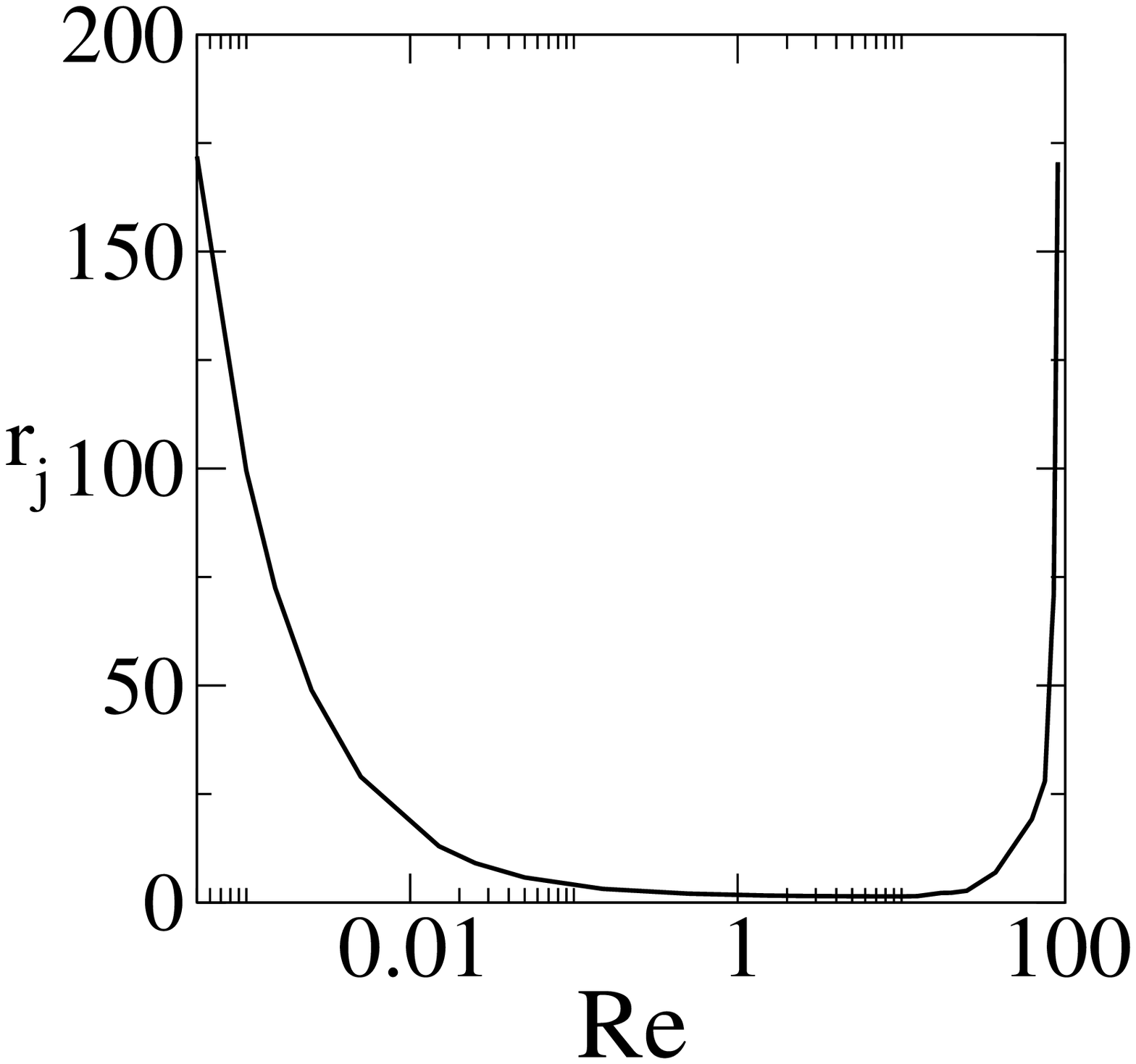}
\includegraphics[width=0.25\textwidth]{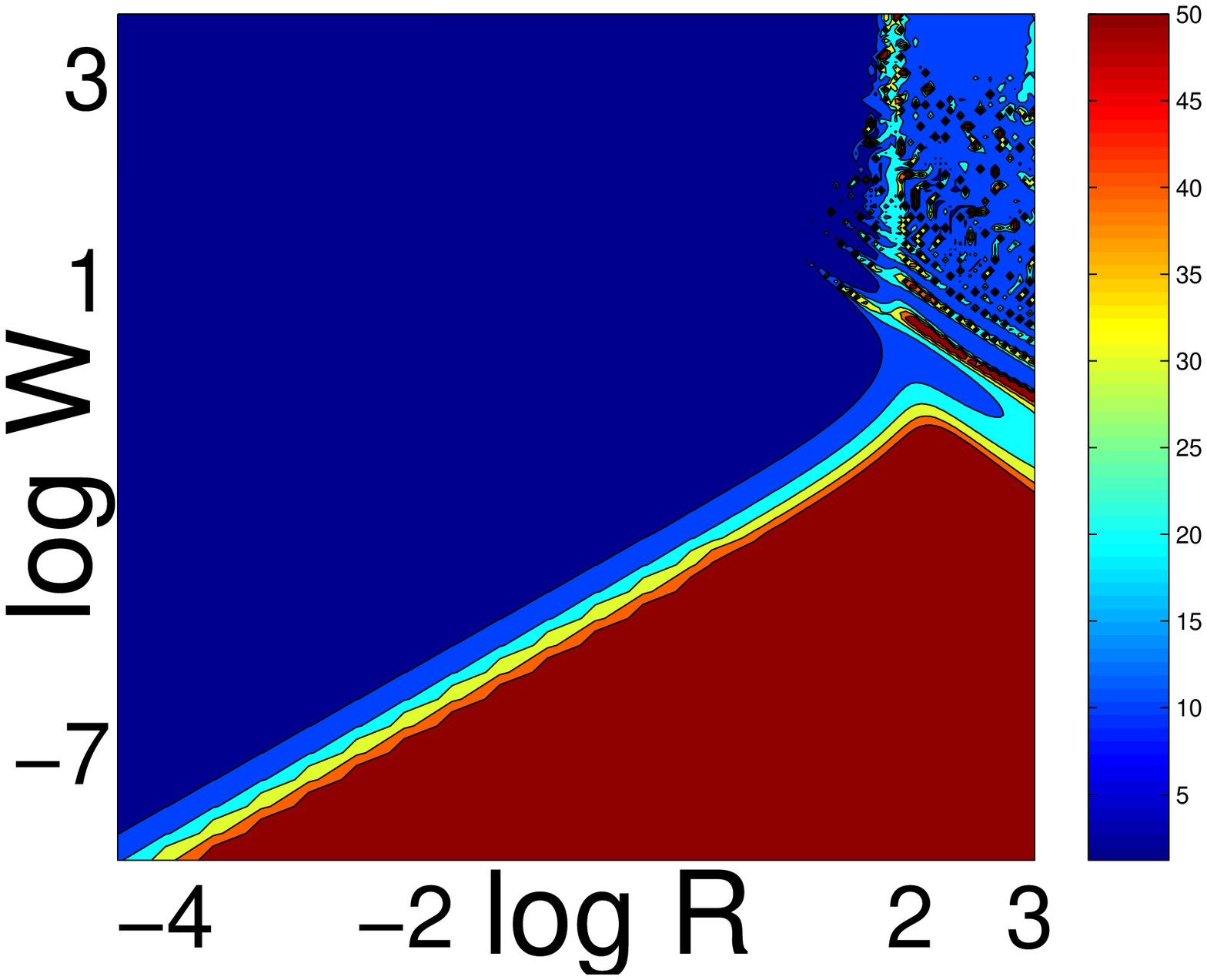}
\caption{(a) The location $r_j$ of the singularity as a function of the inlet
jet radius, expressed here in terms of the Reynolds number. For $R=0.01-90$ 
($u_j=0.02-180$m/s for the case considered) the jump radius is of the order of 
a few microns, as observed in the experiment described below.
(b) Contour plot of jump location in the $R-W$ plane.} 
\label{jumploc}
\end{figure}

We now solve Eq. (\ref{final}) as an initial value problem beginning at 
some location $r_i$ and marching downstream. A fourth-order Runge-Kutta 
algorithm is used. An initial radius $r_i$ somewhat larger than $a$ is chosen, 
where it is assumed that a parabolic profile has been attained. The 
initial
conditions in $h$ and its derivatives are not known exactly for this complicated
problem, and numerical studies are being done to understand the flow in this
vicinity. We have, however, repeated the computations with a variety of initial 
height profiles, and a range $1.2 < r_i < 5$ and $0.1 < h < 1$, and the results 
do not change qualitatively. Typical solutions are shown in figure 
\ref{typical}. At a particular radial location $r = r_j$, there is a 
singularity in the height of the fluid layer. Note that as we approach $r_j$ 
the shallow water equations are no longer valid, even approximately, 
so the present analysis cannot tell us anything about the actual shape close 
to or after the jump. 
The dependence of the jump location on the inlet jet velocity $u_j$
is not monotonic, as seen from figure \ref{jumploc}a. Here the Reynolds number
$R$ is varied by changing $u_j$, with other quantities as in figure
\ref{typical}, so the Weber  
number increases as $R^2$, from $3 \times 10^{-9}$ to $90$. 
For very low $R$ or very high $W$, jumps are unlikely to form within the 
available radius, i.e., inertia and surface tension must be in the right 
balance. The Reynolds and Weber numbers are now varied independent of each
other (figure \ref{jumploc}b). In the region shown in red $r_j > 60$ so jumps 
are not predicted. (A higher cut-off does not
change answers qualitatively.) Blue color indicates
$r_j \sim r_i$, this region merits numerical investigation.
Gravity-free hydraulic jumps may be expected in the region shown by
intermediate color, seen as a relatively narrow linear patch when $R < 100$. 
Here the jump location depends only on the ratio $W/R$. For a given $W$, jumps
exist for over an order of magnitude variation in $R$.
At $R > 100$, undular jumps are seen, which are being investigated further. 

\begin{figure}
\includegraphics[width=0.45\textwidth]{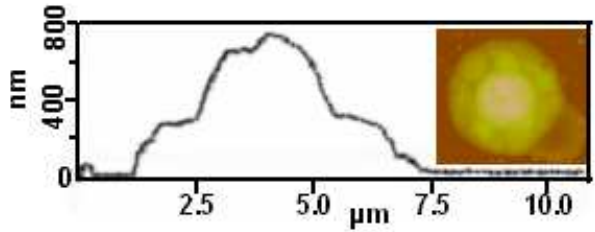}
\caption{AFM image (right) and height profile (left) of a silver blob on 
silicon substrate kept at room temperature.}
\label{blobs}
\end{figure}

\begin{figure}
\includegraphics[width=0.46\textwidth]{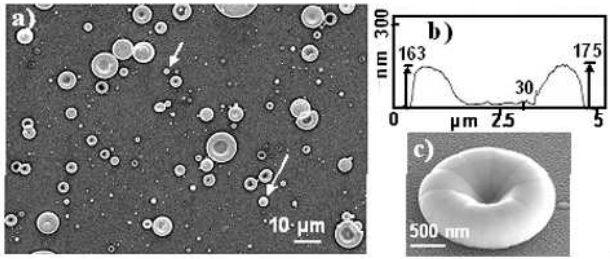}
\caption{Microscopy analysis of metal cups (a) SEM image of femtocups of
silver on a silicon substrate obtained at $E_p=100$ mJ/pulse and $T_s=1173 K$.
A few blobs do exist, as indicated by arrows.
(b) Typical height profile of a femtocup from AFM analysis. (c) Tilted field
emission SEM image of a tin femtocup on silicon.}
\label{egcup}
\end{figure}

We turn now to our experiments, which show a height discontinuity in spreading
drops of molten metal. Since the experimental flow is transient in nature,
a detailed comparison with the theoretical results is not possible, but the jump
radius is in the right range. At larger scales hydraulic jumps are known to
occur even when the incoming flow is in droplets rather than jets \cite{29}. 
A Q-switched frequency tripled Nd:YAG laser 
($\lambda=355$ nm, repetitive frequency, 10 Hz) is focused with 
pulse energy $E_p$ on a rotating metal disc in a vacuum chamber 
($10^{-7}$ torr)
and the resultant plume received at a distance of 4 cm on a clean vertical substrate 
held at a temperature $T_s$, for a duration of 20 min
\cite{kulk}. The resulting metallic structures on the substrate 
are studied by scanning electron microscopy (SEM), atomic force microscopy
(AFM) and energy dispersive X-ray analysis (EDAX). Over most of the range of 
$E_p$ and $T_s$, we expect, and obtain, 
ill-shaped blobs of solidified metals, see figure \ref{blobs}. However, for a
small range of these parameters, there is a strong preference to form 
cup-like structures of outer diameters $\sim 300$nm to $10 \mu$m, with side 
walls $\sim 100$nm high, and capacity $\sim 1$ fL (fig. \ref{egcup}).
The jump diameter is usually about half the total diameter.
Height profiles associated with atomic force micrographs (figure \ref{egcup}b) 
as well as EDAX spectra (not shown) \cite{kulk} confirm that the 
central region is raised from the substrate and contains metal. 
Interestingly, pulsed-laser ablation has been used extensively to produce a 
variety of structures \cite{eglaser}, but femtocups have not been reported 
before, although we notice stray instances of similar structures in other
studies \cite{stray}.

\begin{figure}
\includegraphics[width=0.40\textwidth]{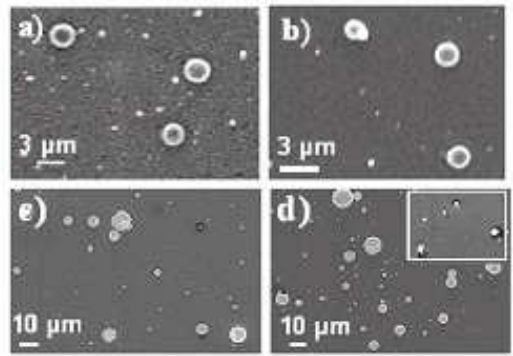}
\caption{SEM images with different metals, laser energy and substrate 
temperature. (a) Cu femtocups on Si. $E_p$=100 mJ/pulse, $T_s = 300$ K; (b) 
Nb on Si, $E_p$=100 mJ/pulse, $T_s=1273$ K; (c) Ag on Si, $E_p=60$ mJ/pulse. 
The number density of well-formed cups is lower. (d) Ag on Si, $E_p=100$
mJ/pulse, $T_s=773$ K much less than $T_m=1234$ K. The cups are not 
well-formed. Inset, $T_s=1273$ K $> T_m$. Only patches are observed.}
\label{various}
\end{figure}

\begin{figure}
\includegraphics[width=0.45\textwidth]{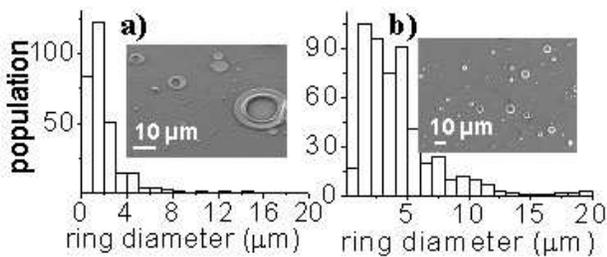}
\caption{Histograms and SEM images of tin femtocups on (a) glass and (b) 
silicon, deposited simultaneously.}
\label{hist}
\end{figure}

This surprising femtocup structure is consistent with the proposed dynamics: 
of a droplet spreading out thinly initially and then undergoing a height 
discontinuity. We obtain femtocups of gold, silver, copper, tin and niobium 
of repeatable statistics on glass, silicon and
graphite (HOPG), see examples in figure \ref{various}a and b, and later in
figure \ref{hist}. 
The solid surface being vertical and the length scales small mean that the effect of 
gravity is negligible. Inertia on the other hand is considerable, since 
velocities are high. We do not have a direct estimate of $u_j$, but we 
may estimate it from earlier measurements in many similar experiments 
\cite{book} to range between 1m/s to 100 m/s. 
Also, the range of $R$ and $W$ over which a 
surface-tension driven hydraulic jump occurs translates to a particular range 
of $E_p$, since laser fluence determines scales and speeds in the incoming jet,
compare figures \ref{egcup}a  and \ref{various}c.
While the substrate is hotter than the metal's
melting point $T_m$, the cup may form initially but cannot solidify and 
liquid flows back into the cup, so the final object is as seen in the inset of 
figure \ref{various}d. 
With $T_s$ far below $T_m$ the tendency to form cups is much reduced (figure
\ref{various}d, probably
because solidification is too rapid for flow to be completed. 
Optimal
conditions are thus $E_p$ ($\sim$ 100 mJ/pulse for silver) and $T_s$ close to but 
below $T_m$. Outside the correct range, blobs form rather than cups.

\begin{figure}
\includegraphics[width=0.40\textwidth]{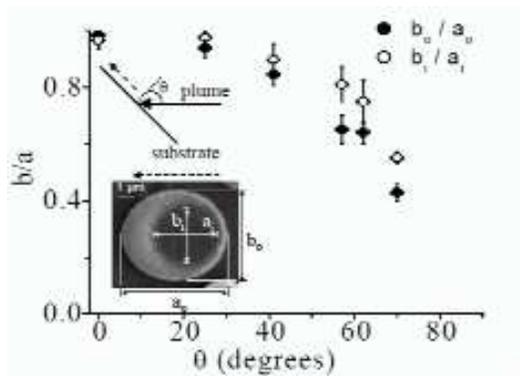}
\caption{Elliptical femtocups with inclined jets. Inset: Sample SEM image of tin
obtained at $\theta=40 \pm 1.5^\circ$, defining $a$ and $b$. The ratio $b/a$
increases with $\theta$, each data point is an average on several cups. The
major axis lies along the direction of maximum flow (dashed arrows).}
\label{angle}
\end{figure}

That the jump is directly 
related to droplet dynamics is confirmed by varying the substrate orientation 
with respect to the incoming jet (see schematic in Fig. \ref{angle}). 
As $\theta$, the inclination of the substrate away from the normal, is 
increased, the structures become increasingly elliptical, especially beyond
$40^\circ$, in accordance with the azimuthal variation of $R$ and $W$.

\begin{figure}
\centering
\includegraphics[width=0.40\textwidth]{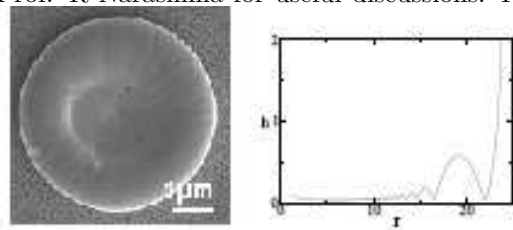} 
\caption{(a) Striations in the structure. (b) An undular hydraulic jump 
at $R=5000$, $W=0.005$.}
\label{floral}
\end{figure}

Since the experiment includes additional complexity in the form of 
solidification, we estimate relative time-scales of jump formation
$t_j$ and solidification of a droplet $t_c$. For the experimental values of 
substrate thickness, $t_c$ ranges from $\sim 3 \times 10^{-4}$s on silicon to 
$\sim 10^{-2}$s on glass (taking into consideration conduction, radiation 
and latent heat), while $t_j \sim r_j/u_j \sim 10^{-6}$s or less.
Contact-line freezing can give rise to an increase in
height in the vicinity, typically amounting to a small percentage of the 
height in the central region. Contrast this to our jump where the height at
the rim is several-fold larger than that in the central region. In spite of
this, and the disparity in
time scales, we cannot rule out a role for local freezing at the contact line
\cite{sonin}. We do notice 
a dependence on the substrate of the size 
distribution of femtocups (figure \ref{hist}) and also some visual differences
in the shape of the femtocup. Our ongoing numerical study, including a 
non-uniform temperature profile and its effects, is therefore aimed at 
a better representation of the experiments. Also being addressed are the
experimental finding of radial striations in the femtocups under certain 
conditions, and the theoretical finding of undular hydraulic jumps (similar to
\cite{bowles} in other conditions) when $R$ is
greater than about $25/W$ (figure \ref{floral}).  

In summary, for the first time, a hydraulic jump solely driven by surface
tension is shown to occur. Experimentally we show evidence for such jumps in 
submicron high inertia droplets of molten metals spreading radially outwards 
on a substrate. The detailed shape in the vicinity of the jump and the transient
problem including the solidification process is being studied numerically.

We are grateful to Prof. G Homsy, Prof. CNR Rao and Prof. R Narasimha for 
useful discussions. RG and NSJ acknowledge support from DRDO (India)
and CSIR (India) respectively.

$*$ To whom correspondence should be addressed, rama@jncasr.ac.in,
kulkarni@jncasr.ac.in

\end{document}